\documentclass[preprint]{revtex4}
\usepackage{graphicx}
\usepackage{amsmath}

\begin{document}

\title{Multi-strange hadrons and the precision extraction of QGP properties in the RHIC-BES domain}

\author{Jussi Auvinen}
\email{jaa49@phy.duke.edu}
\address{Department of Physics, Duke University, Durham, NC 27708, USA}
\author{Krzysztof Redlich}
\address{Institute of Theoretical Physics, University of Wroc\l{}aw, Pl-45204 Wroc\l{}aw, Poland}
\author{Steffen A. Bass}
\address{Department of Physics, Duke University, Durham, NC 27708, USA}

\begin{abstract}
We systematically compare an event-by-event transport+viscous hydrodynamics hybrid model to data from the RHIC beam energy scan using a general Bayesian method.
We demonstrate how the inclusion of multistrange hadron observables affects the outcome of the Bayesian analysis
and conduct an in depth analysis of the viability of $\phi$ and $\Omega$ as probes of the transition region between a deconfined quark-gluon plasma and hadronic phase in heavy ion collisions at higher-end RHIC collision energies.
Utilizing UrQMD to model the final hadronic interactions, we examine the collision rates of $\phi$ and $\Omega$ and the modification to their transverse momentum spectra due to these interactions.
\end{abstract}

\maketitle

\section{Introduction}

Many modern models of relativistic heavy ion collisions employ a so called hybrid approach,
where the hydrodynamical model of quark-gluon plasma evolution is coupled to a hadron transport ``afterburner",
allowing the chemical and kinetic freeze-outs to happen dynamically.
However, this approach introduces an additional parameter to the model;
namely, the switching condition between hydrodynamics and hadron transport.
Typically this switching condition is either a particular value of temperature $T_{SW}$, or energy density $\epsilon_{SW}$, which is close to, but below the transition temperature or energy density.

The justification of this approach comes from the assumption
that both hydrodynamics and hadron transport describe the same system over a range of temperature / energy density values,
and thus the exact value of $T_{SW}$ or $\epsilon_{SW}$ should not matter.
However, to make quantified statements about the actual size of the overlap between the two descriptions,
one needs to identify the experimental observables which probe the transition region.

Multi-strange hadrons $\phi$ and $\Omega$ are potential probes of the transition stage,
as they are produced at the phase boundary during hadronization of the
quark-gluon plasma and exhibit small scattering cross section in the hadronic phase \cite{Shor:1984ui}. In this Article, we perform a detailed comparison of an event-by-event transport+viscous hydrodynamics hybrid model \cite{Karpenko:2015xea}
to $\phi$ and $\Omega$ data from the RHIC beam energy scan \cite{Abelev:2008aa,Aggarwal:2010ig,Adamczyk:2015lvo}.

\section{Hybrid model}
\label{sec:model}

In the hybrid approach, the heavy ion collision is modeled in three separate phases.

The initial pre-equilibrium phase is simulated with UrQMD hadron+strings cascade \cite{Bass:1998ca,Bleicher:1999xi}.
The hydrodynamical evolution starts after the two colliding nuclei have passed through each other: $\tau_0 \geq 2R_{\text{nucleus}}/\sqrt{\gamma_{CM}^2-1}$.
At this point, the particle properties such as energy and baryon number are converted to densities
using 3D Gaussians with ``smearing'' parameters $R_{\text{trans}}$, $R_{\text{long}}$ , each equal to $\sqrt{2}$ times the respective Gaussian width parameter $\sigma_{\text{trans}}$, $\sigma_{\text{long}}$.

In the local equilibrium phase, the system is evolved according to 3+1D viscous hydrodynamics \cite{Karpenko:2013wva},
with viscosity parameter $\eta/s$ kept constant during the full evolution.
At the lower collision energies, the equation of state needs to include the effects of nonzero net-baryon density.
For this purpose, a chiral model equation of state \cite{Steinheimer:2010ib} is utilized in this investigation.

Finally, the transition from hydrodynamical evolution back to hadron transport (``particlization'') happens when 
energy density $\epsilon$ drops below the chosen energy density value $\epsilon_{SW}$.
A hypersurface with constant energy density is constructed \cite{Huovinen:2012is} and
particles are sampled from this hypersurface according to the Cooper-Frye formula and propagated further using UrQMD.
Both chemical and kinetic freeze-out thus happen dynamically.

As all $\phi$s will decay before the end of the simulation,
we output the full interaction histories of the afterburner hadron cascades for each event.
We search these histories for $\phi$s,
and label the ones which have both their decay products surviving to the end of simulation as ``detectable".
The decay products which have experienced only soft scatterings are included in survivors.

\section{Statistical analysis and simulation setup}
\label{sec:stats}

The optimal input parameter values are determined using Bayesian analysis similar to Refs. 
\cite{Novak:2013bqa,Bernhard:2015hxa}.
The posterior distribution is sampled with Markov chain Monte Carlo (MCMC) method.
We perform thousands of random walks in input parameter space,
where step proposals are accepted or rejected based on a relative likelihood.
However, it is not feasible to run the full hybrid model simulation for each evaluation of the likelihood function.
To circumvent this problem, we use Gaussian processes to emulate the simulation output, based on
$\approx 100$ training points, which are samples of the input parameter space. The Latin hypercube method is used
to achieve close to uniform distribution of training points on all 5 parameter dimensions.

\section{Results}
\label{sec:results}

\begin{figure}
\includegraphics[width=5cm]{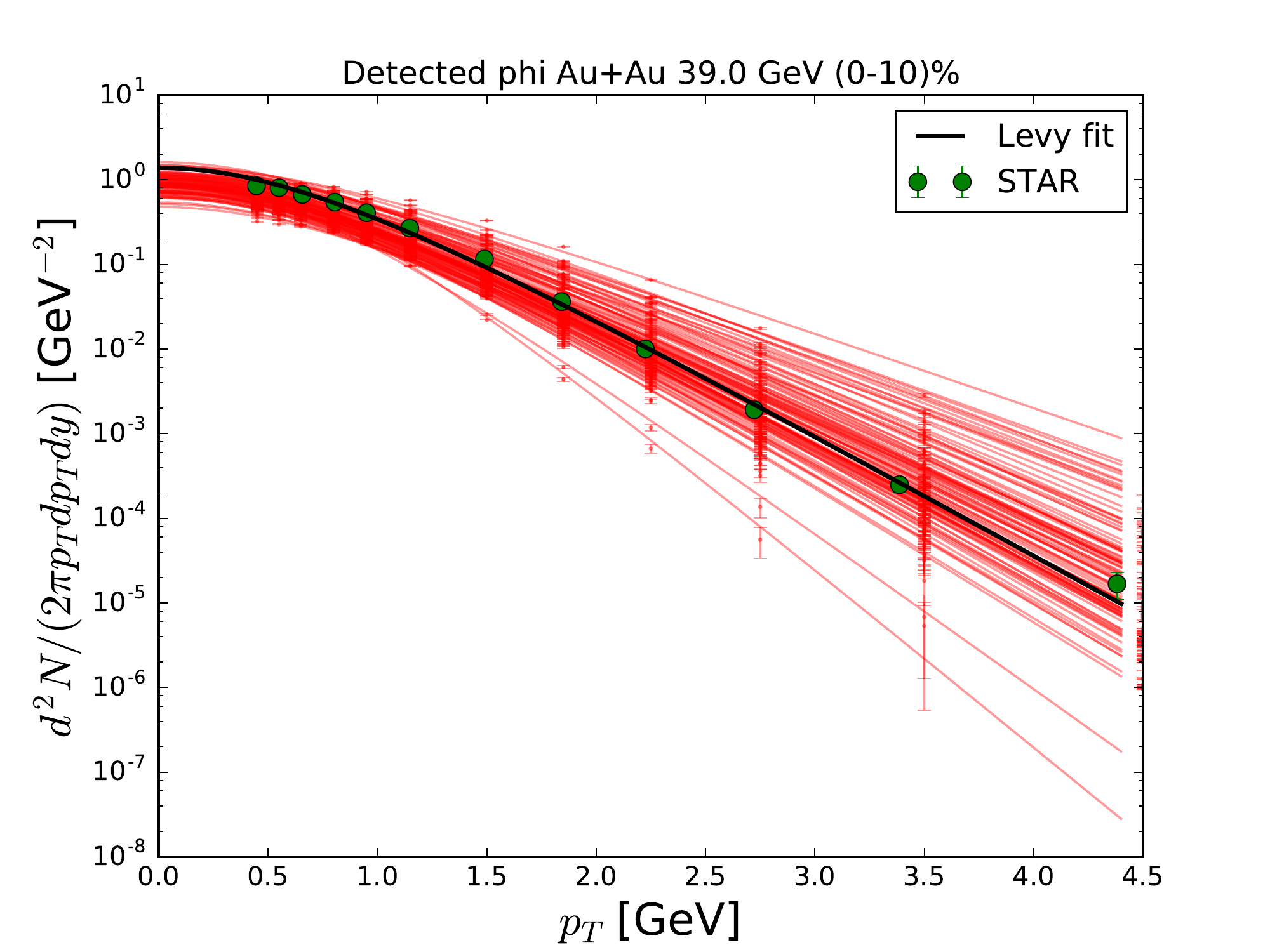}
\includegraphics[width=5cm]{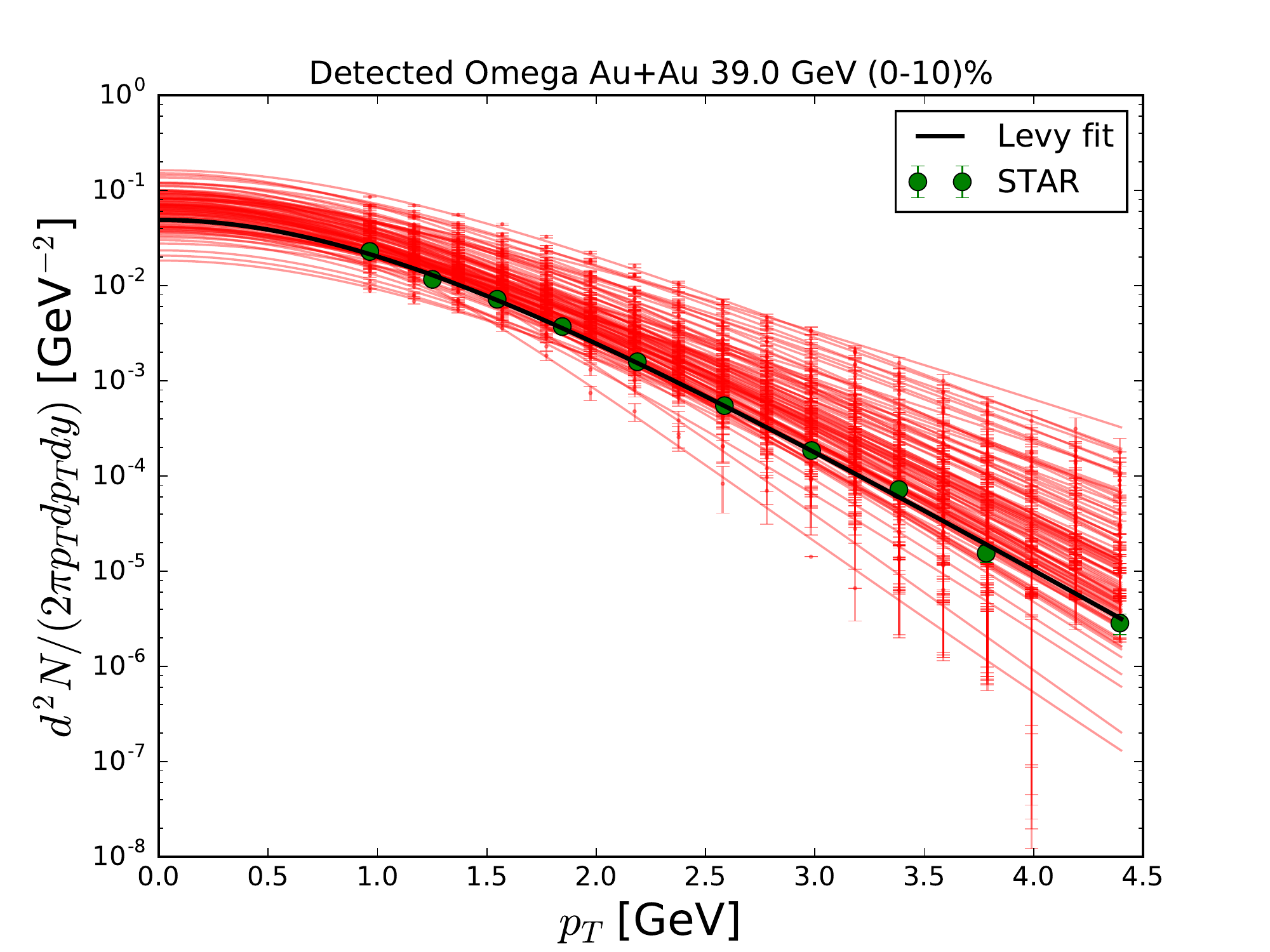}
\includegraphics[width=5cm]{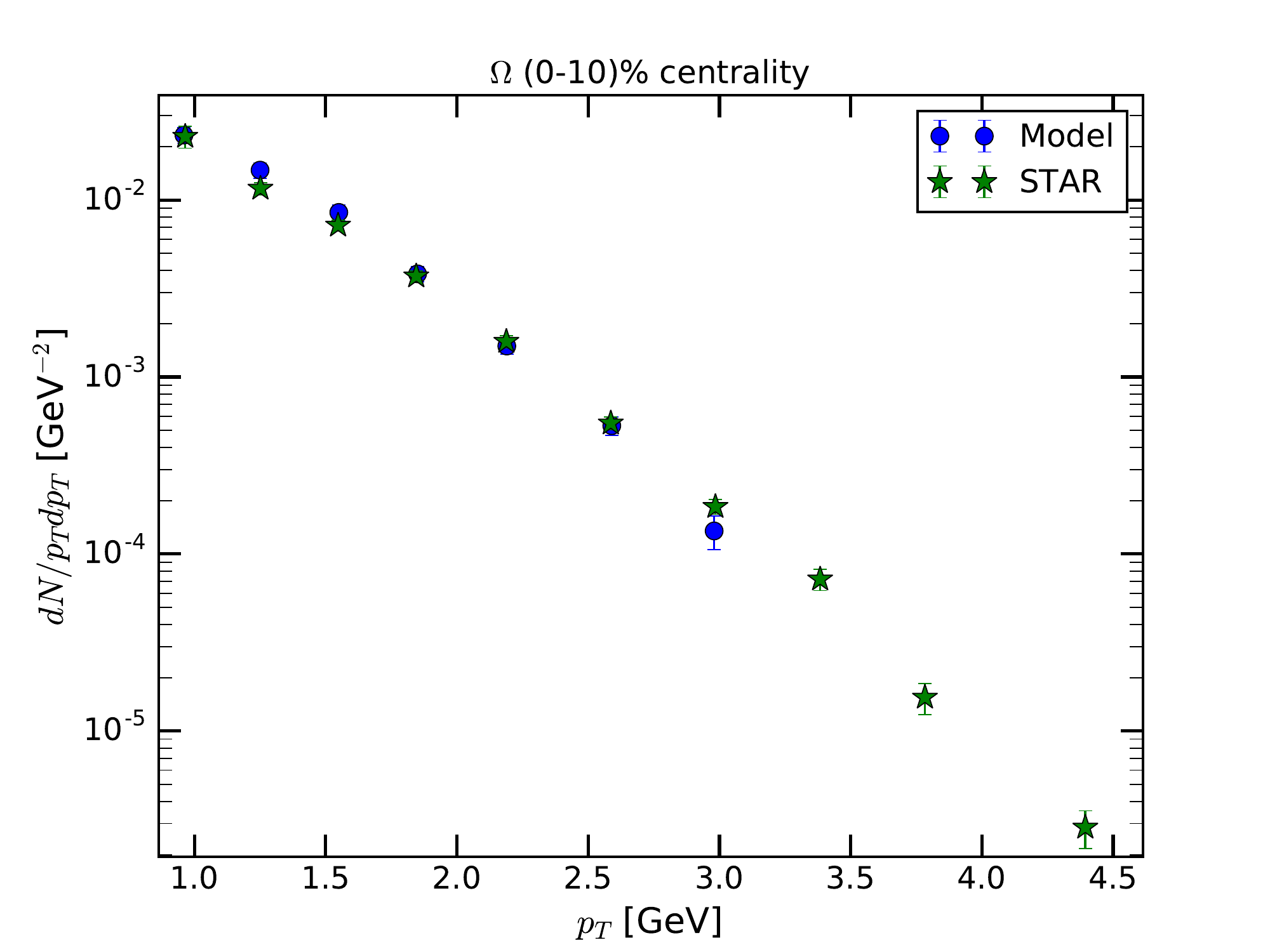}
\caption{Transverse momentum spectra of $\phi$ (left) and $\Omega$ (middle) at $\sqrt{s_{NN}}=39$ GeV.
Curves represent Levy fits on model or STAR data points.
Right: Transverse momentum spectrum of $\Omega$ at $\sqrt{s_{NN}}=39$ GeV using the median values
from statistical analysis.
STAR data from \cite{Adamczyk:2015lvo}.}
\label{fig:ptspectra}
\end{figure}

Figure \ref{fig:ptspectra} demonstrates the variance in transverse momentum distributions of $\phi$s and $\Omega$s
over the training points for $\sqrt{s_{NN}}=39$ GeV. 
To eliminate some possible sources of discrepancies between simulation and experimental results,
we follow the experimental method \cite{Abelev:2008aa} and do also Levy fits on the transverse momentum spectra points
which are within the $p_T$ range reported by the experiments.
We see that $\phi$ yield at low $p_T$ is underpredicted by most of the input parameter combinations. 
This suggests that our criterion for $\phi$ meson detection is too stringent compared to the experimental reconstruction methods.

To verify the result of the statistical analysis,
a full simulation was performed using the median values from the posterior distributions of the input parameters.
The resulting $\Omega$ transverse momentum distributions for $\sqrt{s_{NN}}=$  39 GeV is shown in the right frame of Fig.~\ref{fig:ptspectra}.
Although only the lowest transverse momentum bin was used for the calibration of model parameters,
the experimental data is very well reproduced for the whole $p_T$ spectrum.

In addition of the yields, we can also calculate the mean $p_T$ from the particle spectra.
As we are interested in using $\phi$ and $\Omega$ as probes of the phase transition region,
we need to quantify the change on $\langle p_T \rangle$ during the hadron gas phase.
We find the detectable $\phi$ mean transverse momentum to be about 20\% larger than the value at the hypersurface. The increase is slightly smaller, $\approx 15$\%, when including ``undetectable" $\phi$s. This equals the change in $\langle p_T \rangle$ seen for $\Omega$.

\begin{figure}
\includegraphics[width=5cm]{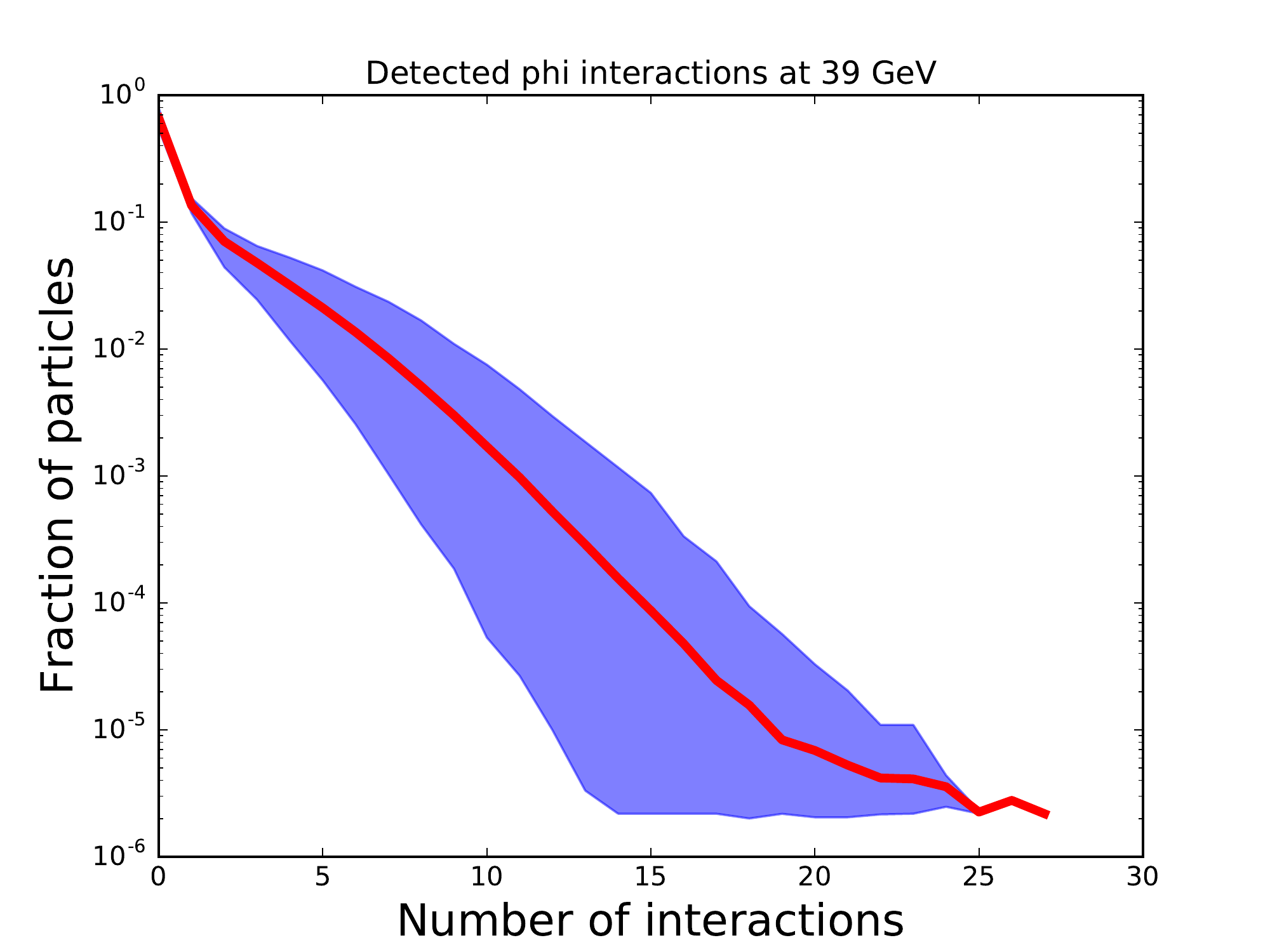}
\includegraphics[width=5cm]{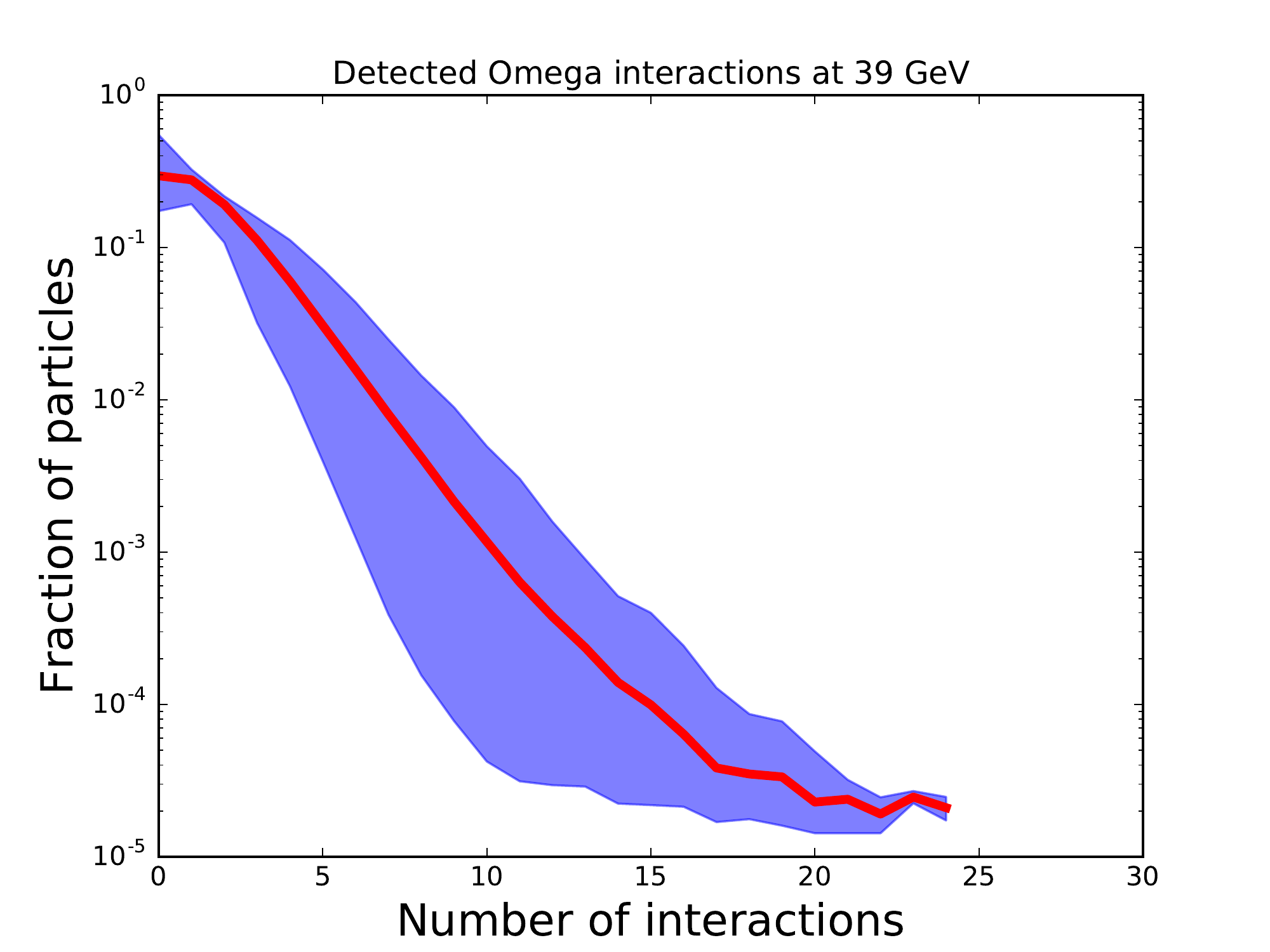}
\includegraphics[width=5cm]{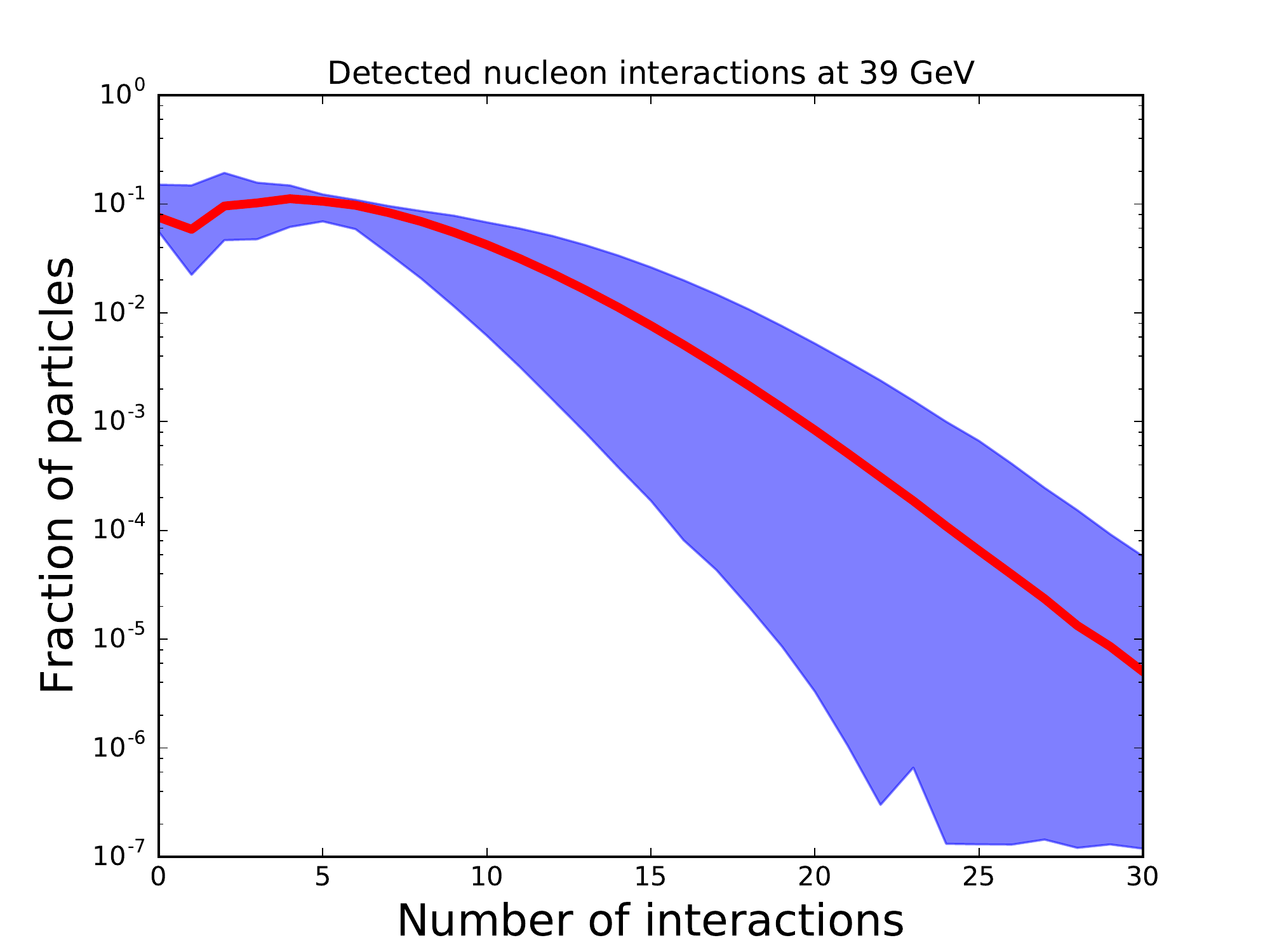}
\caption{Distribution of average number of interactions for $\phi$, $\Omega$ and $N$ at $\sqrt{s_{NN}}=39$ GeV.
Red line represents the mean over all training points.
Blue band represents the full range of values over all training points.}
\label{fig:interactions}
\end{figure}

We also investigate the average number of interactions $\phi$ and $\Omega$ experience in the hadronic matter (Fig.~\ref{fig:interactions}).
The analysis verifies that detectable $\phi$ mesons have hardly any interactions,
and $\Omega$ baryons are likely to have only a few, compared to the 5-10 interactions of a typical nucleon.

\begin{figure}
\includegraphics[width=5cm]{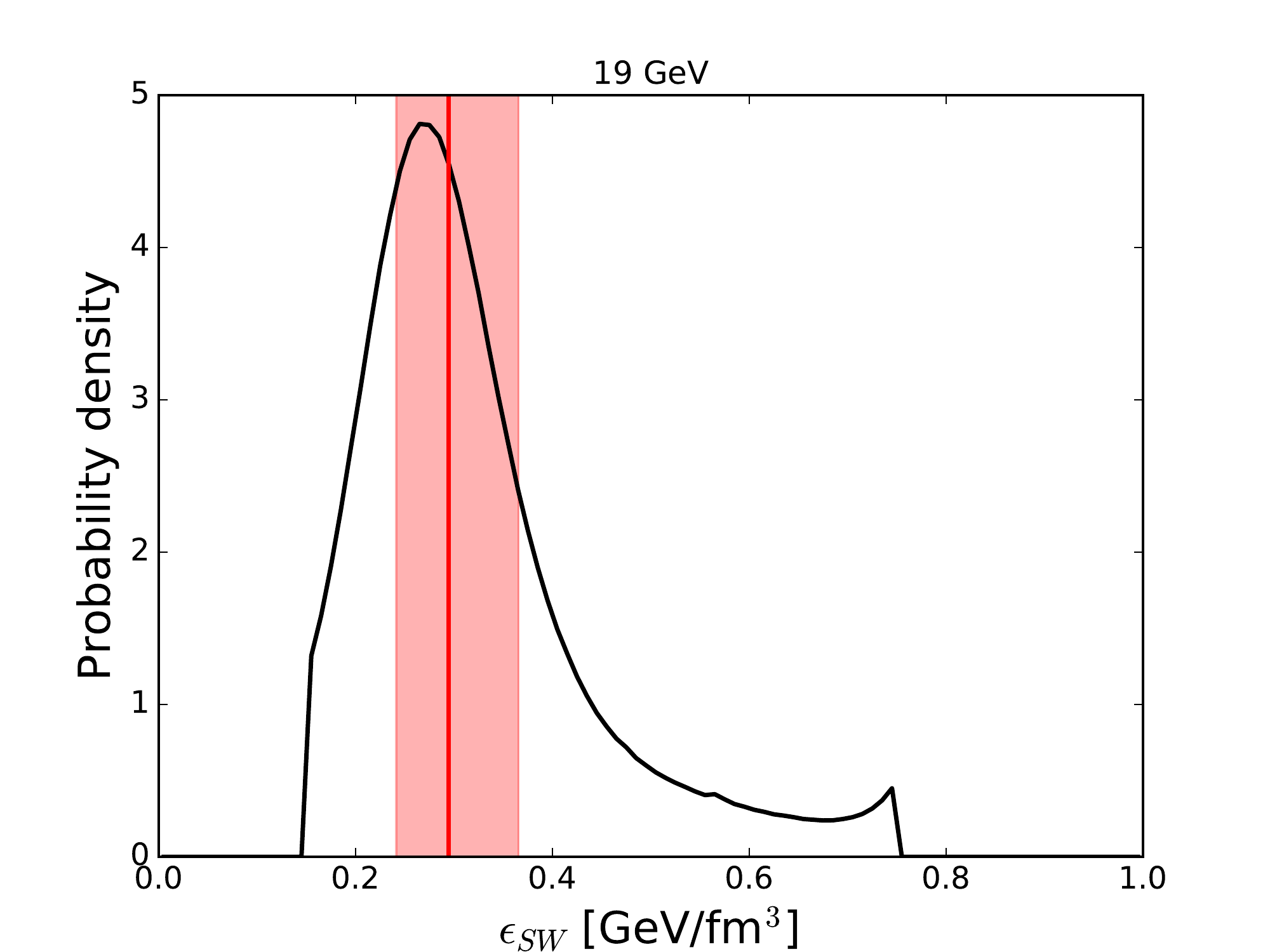}
\includegraphics[width=5cm]{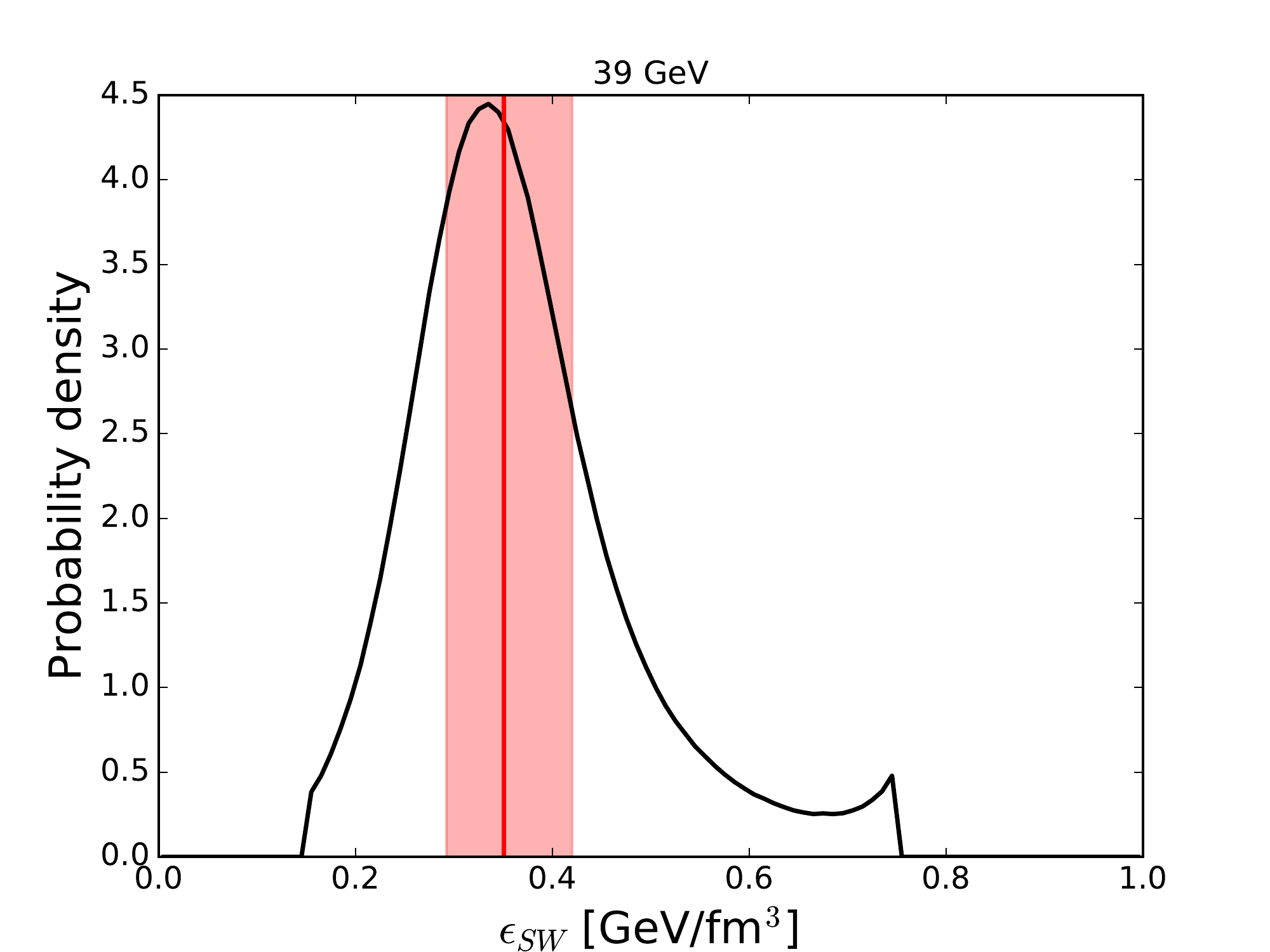}
\includegraphics[width=5cm]{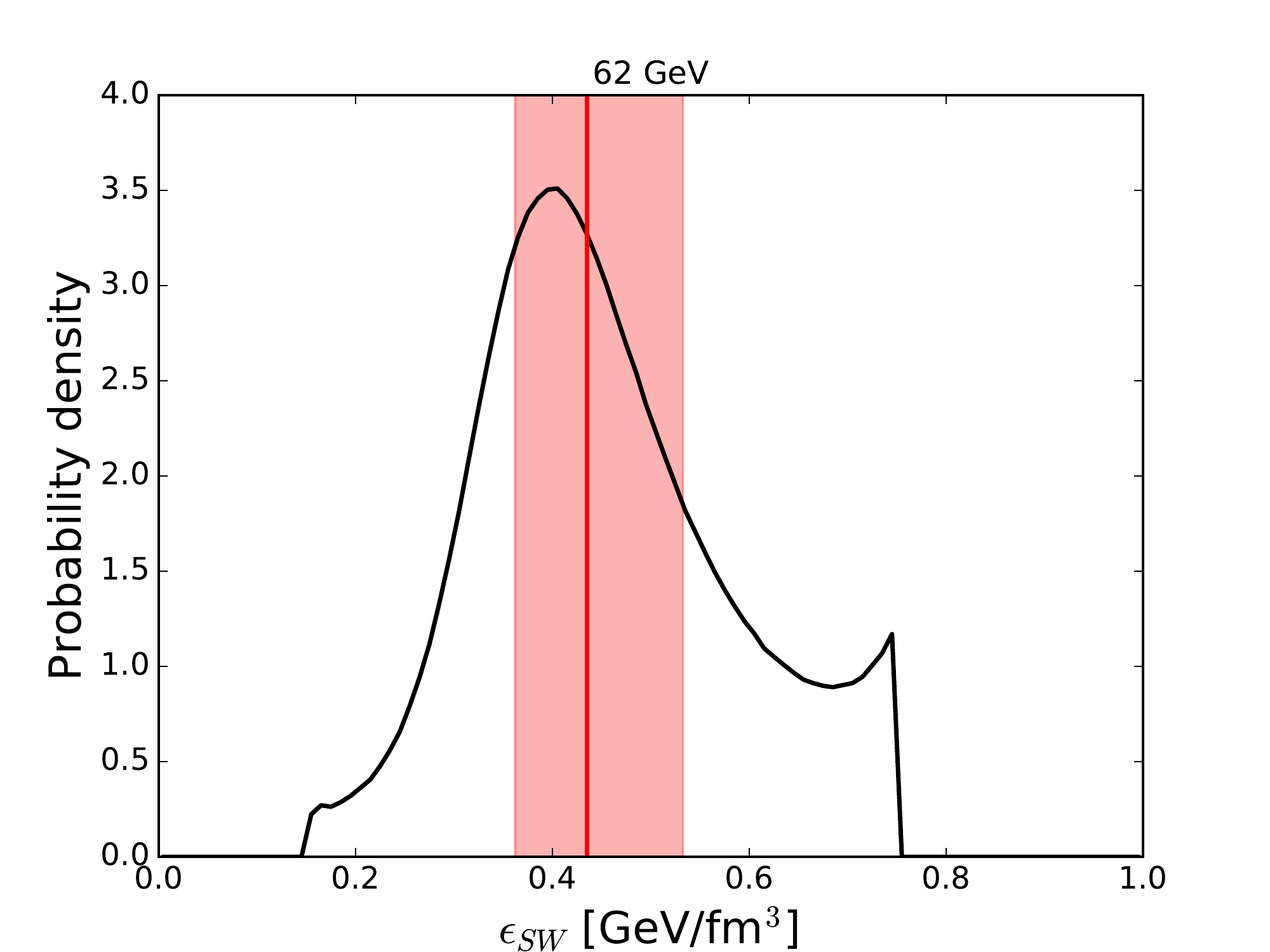}
\caption{Posterior probability distributions of the switching energy density $\epsilon_{SW}$ for different collision energies:
19.6 GeV (left), 39 GeV (middle), and 62.4 GeV (right).
Colored bands and median lines correspond to respective boxes and median lines in the right frame of Fig.~\ref{fig:bayesomegabox}.
}
\label{fig:bayesomegadistr}
\end{figure}

\begin{figure}
\centering
\includegraphics[width=6cm]{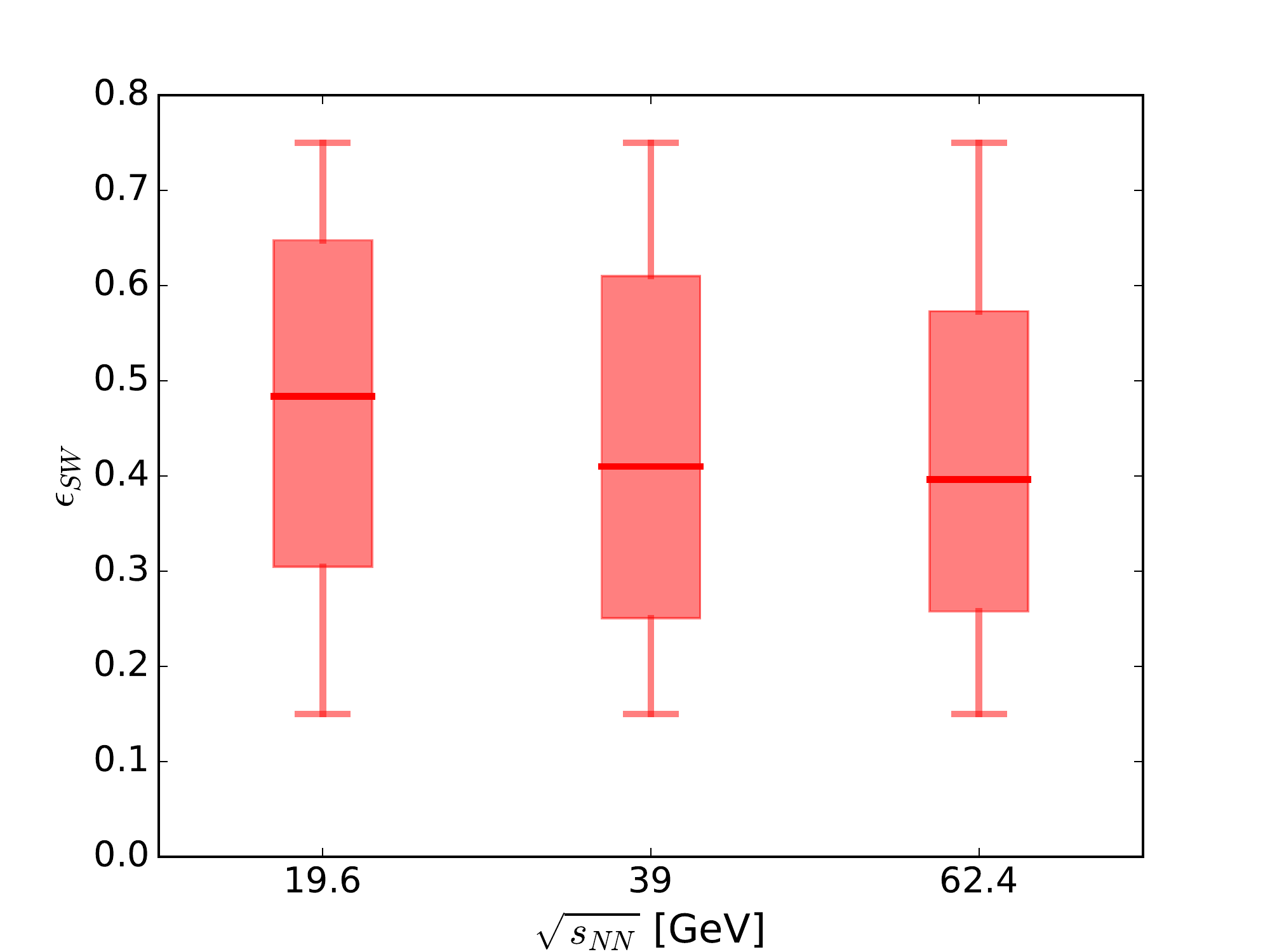}
\includegraphics[width=6cm]{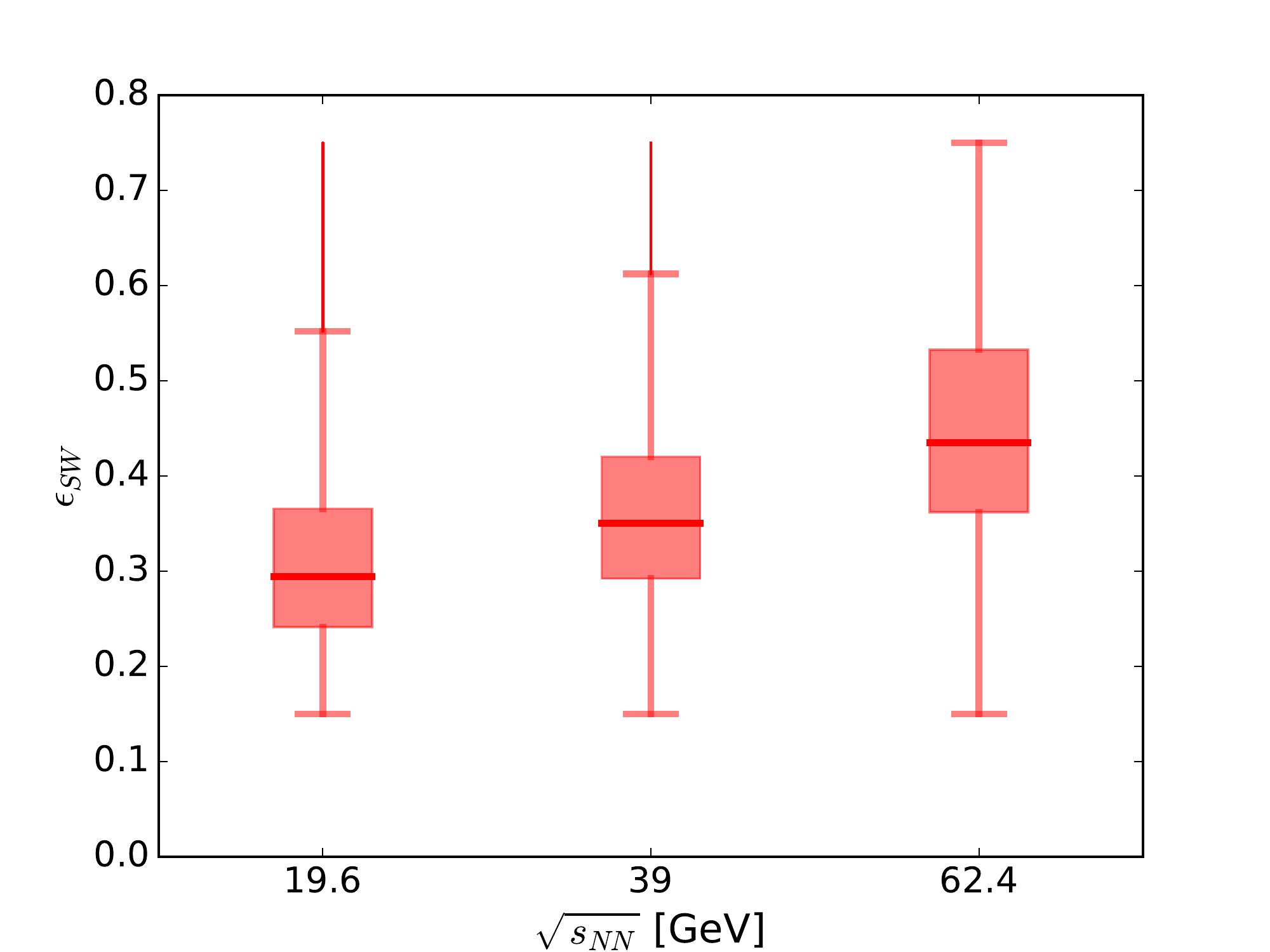}
\caption{Box-whisker plots of the Bayesian posterior distribution of the switching energy density $\epsilon_{SW}$ vs. $\sqrt{s_{NN}}$.
Left: without $\Omega$ yields.
Right: with $\Omega$ yields.
Boxes and median lines of the right-side figure correspond to respective colored bands and median lines in Fig.~\ref{fig:bayesomegadistr}.}
\label{fig:bayesomegabox}
\end{figure}

To check the effect of multistrange hadron observables on the posterior distributions,
the statistical analysis was performed with and without $\Omega$ yield data for $\sqrt{s_{NN}}=$ 19.6, 39, and 62.4 GeV.
As seen in Figs.~\ref{fig:bayesomegadistr} and \ref{fig:bayesomegabox},
the analysis without $\Omega$ yields was not able to produce any constraints on switching energy density value,
whereas the analysis including the $\Omega$ provide clear peak regions in probability distributions,
revealing a visible dependence on the collision energy.

\section{Summary}
\label{sec:summary}
We have performed a detailed investigation of multistrange hadrons in relativistic heavy ion collisions,
focusing on collision energies $\sqrt{s_{NN}}=$ 39 and 62.4 GeV.

Based on their low number of interactions and the relatively small changes on their mean transverse momentum during the hadron gas phase,
we find the assumption of $\phi$ and $\Omega$ as clear probes of the phase transition region justified.
Performing a Bayesian analysis with and without $\Omega$ yield data has demonstrated the large effect the inclusion of multistrange hadron data can have on model-to-data comparisons when determining the ``true'' parameter values.

\section{Acknowledgements}

We thank Jonah E. Bernhard for useful discussions and providing a Python code package for the statistical analysis, and Iurii Karpenko for providing the 3+1D viscous hydrodynamics hybrid model code.
The authors acknowledge support by the U.S. Department of Energy Grant no.~DE-FG02-05ER41367.
K. R. also acknowledges partial support of the Polish Science Center (NCN) under Maestro grant DEC-2013/10/A/ST2/00106.

\section*{References}
\bibliography{sqm2016}

\end{document}